# Positionality-Weighted Aggregation Methods for Cumulative Voting

Takeshi Kato[1], Yasuhiro Asa[1] & Misa Owa[1]

[1] Hitachi Kyoto University Laboratory, Open Innovation Institute, Kyoto University, Kyoto, Japan

Correspondence: Takeshi Kato, Hitachi Kyoto University Laboratory, Open Innovation Institute, Kyoto University, Kyoto 606-8501, Japan.

**Abstract**

Respecting minority opinions is vital in solving social problems. However, minority opinions are often ignored in general majority rules. To build consensus on pluralistic values and make social choices that consider minority opinions, we propose aggregation methods that give weighting to the minority's positionality on cardinal cumulative voting. Based on quadratic and linear voting, we formulated three weighted aggregation methods that differ in the ratio of votes to cumulative points and the weighting of the minority to all members, and assuming that the distributions of votes follow normal distributions, we calculated the frequency distributions of the aggregation results. We found that minority opinions are more likely to be reflected proportionately to the average of the distribution in two of the above three methods. This implies that Sen and Gotoh's idea of considering the social position of unfortunate people on ordinal ranking in the welfare economics, was illustrated by weighting the minority's positionality on cardinal voting. In addition, it is possible to visualize the number and positionality of the minority from the analysis of the aggregation results. These results will be useful to promote mutual understanding between the majority and minority by interactively visualizing the contents of the proposed aggregation methods in the consensus-building process. With the further development of information technology, the consensus building based on big data will be necessary. We recommend the use of our proposed aggregation methods to make social choices for pluralistic values such as social, environmental, and economic.

## 1. Introduction

Disparities in wealth, inequalities in income, environmental pollution and cultural collapse amongst others, have become social problems in various countries, regions, and communities. To solve these social problems, it is not only necessary to consider economic values, but also consider pluralistic values related to human beings, society, environment, and culture (e.g. Epstein & Buhovac, 2014). To facilitate the development of a sustainable and equitable society, social choices ought to be made concurrent with building consensus among stakeholders who usually have varying values. The usual social choice is decided through a voting process and the aggregation of diverse opinions (e.g. Craven, 1992).

However, voting to opt for the appropriate collective decisions presents a fundamental challenge. Due to Condorcet's paradox (Condorcet, Sommerlad, & McLean, 1989) and Arrow's impossibility theorem (Arrow, 1950), it is theoretically and practically difficult to aggregate various opinions. Various voting methods have been devised by mathematicians, political scientists, economists, among other specialists. These include the majority rule, score voting based on Borda count (Borda, 1784), and cumulative voting based on total number of points (Kenny, 1976; FairVote, 2006). Nonetheless, standard criteria to select a voting method for the preferred social choice is indefinite (e.g. Sakai, 2013; Pacuit, 2019).

Quadratic voting, a recently developed voting method, is a variant of cardinal cumulative voting (Lally & Weyl, 2018; Posner & Weyl, 2018). In quadratic voting, the voter uses points that are evenly distributed among voters who will in turn express their degree of preference. The decided points are then cumulatively assigned to the most favored choice. The term 'quadratic' means the number of votes that could be purchased is a square root function of a number of points based on the economic pricing rule (i.e., the voter should pay the cost to change the cardinal choice because this specific change boosts the marginal cost of all members (Tideman & Plassmann, 2017)).

Opinion surveys have shown that quadratic voting-based surveys could be used to accurately measure respondent preferences when compared to the de facto standard Likert survey (Quarfoot et al., 2017). In a Likert survey, aggregated results from responses for a discrete index axis (for example, 7th quantile) have a W-shaped frequency distribution which is extremely biased. In comparison, in a survey based on quadratic voting, the aggregated result for a more subdivided and substantially continuous index axis (for example, 20th quantile) has a bell-shaped quasi-normal distribution. This aggregated result is considered to closely approximate the true distribution of respondents' preferences.

In welfare economics, that is a branch of economics to evaluate social welfare, the concepts of justice and equality are discussed from a viewpoint different from opinion surveys. In Rawls' theory of justice, "original position", in which the peculiarity of an individual's name is obscured by a "veil of ignorance", is the premise for making a fair social choice (Rawls, 1971). In Rawls' view, social positions related to gender, race, and culture are considered to be included in "the general facts of human society" or "generality" (Rawls, 1955).

On the other hand, Sen's idea of justice, an extension of Rawls' theory, assigns asymmetric weights to the "social position" of vulnerable groups people due their race, gender, disability, etc. This social position broadly represents a shared historical, social, cultural or personal characteristic. The qualitative difference between the position of such groups and other groups is the subject of social concern. Additionally, positional parameters in this context are reflectively and continuously updated through "trans-positional assessment" (Sen, 1999; Sen, 2002). Gotoh's literature (Gotoh, 2014) details the controversy between the "social positions" of Rawls and Sen.

A position-conscious choice procedure is a social-choice method that reflects Sen's position idea (Gotoh, 2015; Gotoh, 2017). In this procedure, priority is given to the rank of a victim group on an ordinal scale based on specific values (priority condition), the rank agreed among victim groups regarding different values is defined as the social ranking (group Pareto condition), and each of the victim groups does not interfere the other group's ranking regarding other values with this process (refrain condition). Therefore, social choices are made in based on the vulnerability of different groups. However, in the event of an unexpected case or the appearance of new victims, this choice procedure will be corrected in a consensus-building process.

Here, cardinal cumulative voting, including quadratic voting, has the advantage of obtaining aggregate results that represent the true voters opinions' distribution. These voters are also stakeholders. Furthermore, an analysis of voting data can yield the votes of voters who have strong preference for specific values through the comparison of the back-calculation result to 1p1v (1 person 1 vote) and the cumulative voting result, and can extract minority's votes. Moreover, the aggregation results can be changed based on weights that take into consideration the minority's position based on the idea of Sen and Gotoh. Although it is not possible to know the content of the social position of the minority from the voting data alone, the aggregation weight discloses the position of the cardinal value (positionality) supported by the minority on the voting.

We aspire to make social choices that reflect the opinions of the minority, e.g., unfortunate people. To achieve this purpose, we propose new aggregation methods that give weight to the minority's positionality by incorporating the position-conscious concept of Sen and Gotoh into cumulative voting. In investigating the properties of these methods, and assuming a normal distribution as the true opinion distribution of voters, quadratic and ordinary point voting (so-called linear voting) procedures are then used as cumulative voting methods for pluralistic values. We then discuss how the aggregation results change when aggregation methods give weight to the minority's positionality. Specifically, we formulate our aggregation methods in Chapter 2, present the aggregation results in Chapter 3, discuss these methods' characteristics and use in the consensus-building process in Chapter 4, and conclude the paper in Chapter 5.

**2. Our Proposed Aggregation Methods**

We formulated our aggregation methods to give weight to the minority's positionality during cardinal cumulative voting. There is an assumption that for pluralistic values, the votes of the majority are evenly distributed, the minority cumulatively votes for a particular monistic value, the frequency distributions of the aggregation results of the majority and minority follow normal distributions, and both distributions have a reproductive property (the probability distribution for the sum of stochastic variable of the majority and the stochastic variable of the minority follows normal distribution, i.e., these distributions can be composed or decomposed).

Figure 1 shows the frequency distributions of cumulative voting for a certain monistic value. The black line shows the positionality-weighted distribution of all members, and the green line shows the non-weighted original distribution of all members. The blue line and the orange line show the original distribution of the majority and the minority each other. The dark-red line shows the positionality-weighted distribution of the minority.

The black line is obtained from the raw distribution of the weighted aggregation result. The green line is obtained by analyzing the raw distribution (similar to 1p1v conversion). The blue line and the orange line are obtained by decomposing the original distribution of all members. The dark-red line is obtained by weighting the minority's positionality for the original distribution of the minority. In Fig. 1, the black-line distribution is composed of the blue-line distribution, the orange-line distribution (right side), and the dark-red-line distribution (left side). The formulation conditions are listed in itemized form below.

- $n$ members distribute their votes for $k$-dimensional pluralistic values.
- The original frequency distribution of all members $n$ with '1 person same vote' (similar to 1p1v) for a certain one-dimensional coordinate follows the normal distribution $N(\mu_o, \sigma_o^2)$.
- Points $p$ per person are distributed for cumulative voting.
- The number of minority voters with strong preference is $m$ out of $n$, and the number of majority voters with weak preference is $n - m$.
- Each of the majority voters votes $a$ for a certain one dimension by evenly distributing the $p$ for $k$ dimensions, and its distribution follows the original normal distribution $N(\mu_o, \sigma_o^2)$.
- Each of the minority voters votes $b$ by cumulatively concentrating the $p$ for a certain one dimension, and its distribution follows the normal distribution $N(\mu_m, \sigma_m^2)$.
- Positionality weight $w_p$ is given to the minority. However, when $k = 1$, there is no difference in the number of votes between the majority and minority ($a = b$), and the minority cannot be recognized, so $w_p = 1$.
- The distribution of the positionality-weighted aggregation that combines the distributions of the majority and minority is $N(\mu, \sigma^2)$.
- For convenience of calculation, the distribution $N(2\mu_o - \mu_m, \sigma_m^2)$ of the mirror image of the minority $m$ is assumed.

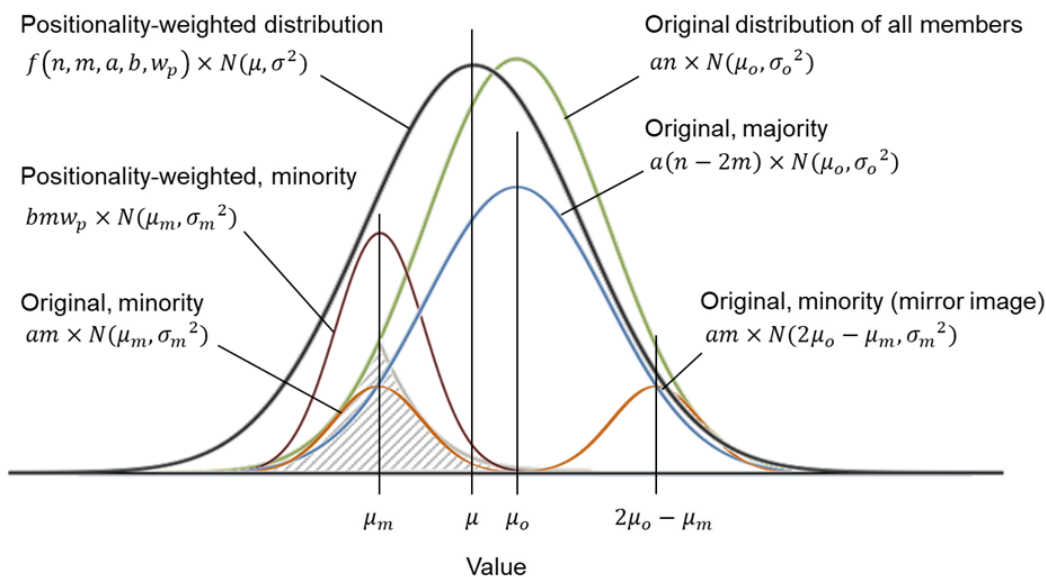

Figure 1. Frequency distributions of all members; majority and minority

Under these conditions, the mean $\mu$ of the distribution $N(\mu, \sigma^2)$ (black line in Fig. 1) of the positionality-weighted aggregation is calculated from the weighted mean of the combination of the distributions (blue, dark-red, and orange lines on the right side in Fig. 1) and is expressed using the following equation.

$$\mu = \frac{a(n-2m)\mu_o + am(2\mu_o - \mu_m) + bmw_p\mu_m}{a(n-2m) + am + bmw_p} = \mu_o \frac{1 + \left(\frac{b}{a}w_p - 1\right)\frac{m}{n}\frac{\mu_m}{\mu_o}}{1 + \left(\frac{b}{a}w_p - 1\right)\frac{m}{n}}$$

(1)

The maximum number of the minority voters at $\mu_m$ in the original frequency distribution (green line in Fig. 1) is calculated from the cumulative distribution function of the normal distribution (the gray-shaded area in Fig. 1), as shown in Eq. (2).

$$\mu_m < \mu_o$$

$$\frac{m}{n} < 2 \cdot \frac{1}{2}\left(1 + \mathrm{erf}\frac{\mu_m - \mu_o}{\sqrt{2\sigma_o^2}}\right) = 1 + \mathrm{erf}\frac{\frac{\mu_m}{\mu_o} - 1}{\sqrt{2}\frac{\sigma_o}{\mu_o}}$$

$$\mu_m \geq \mu_o$$

$$\frac{m}{n} < 1 - 2 \cdot \frac{1}{2}\left(1 + \mathrm{erf}\frac{\mu_m - \mu_o}{\sqrt{2\sigma_o^2}}\right) = 1 - \mathrm{erf}\frac{\frac{\mu_m}{\mu_o} - 1}{\sqrt{2}\frac{\sigma_o}{\mu_o}}$$

(2)

In Eqs. (1) and (2), the five methods, i.e., quadratic-non-weighted (Q-NW), quadratic-square-root weighted (Q-SW), quadratic-linear weighted (Q-LW), linear-linear weighted (L-LW), and linear-non-weighted (L-NW), are formulated 1 by changing $a$ and $b$ in cumulative voting and $w_p$ in aggregation. Quadratic and linear voting are assumed as the cumulative voting methods, and the majority/minority ratio $m/n$ and its square root $\sqrt{m/n}$ are assumed as weights. Q-NW is a usual quadratic voting method, L-NW is a usual linear voting method, and Q-SW, Q-LW, and L-LW are new proposed methods. This is shown in table 1 below.

Table 1. Usual and proposed aggregation methods

| Method | Q-NW (usual) | Q-SW (proposed) | Q-LW (proposed) | L-LW (proposed) | L-NW (usual) |
|---|---|---|---|---|---|
| Voting method | Quadratic | Quadratic | Quadratic | Linear | Linear |
| Aggregation method | Non weighted | Square root weighted | Linear weighted | Linear weighted | Non weighted |
| Majority voting $a$ | $\sqrt{\frac{p}{k}}$ | $\sqrt{\frac{p}{k}}$ | $\sqrt{\frac{p}{k}}$ | $\frac{p}{k}$ | $\frac{p}{k}$ |
| Minority voting $b$ | $\sqrt{p}$ | $\sqrt{p}$ | $\sqrt{p}$ | $p$ | $p$ |
| Positionality weight $w_p$ | 1 | $\sqrt{\frac{m}{n}}$ | $\frac{m}{n}$ | $\frac{m}{n}$ | 1 |

The ratio of the mean $\mu$ of the aggregation to the mean of the original normal distribution $\mu_o$ ($\mu/\mu_o$) for Q-NW, Q-SW, Q-LW, L-LW, and L-NW, is calculated using Eqs. (3) to (7), respectively.

Q-NW:

$$\frac{\mu}{\mu_o} = \frac{1 + (\sqrt{k} - 1)\frac{m}{n}\frac{\mu_m}{\mu_o}}{1 + (\sqrt{k} - 1)\frac{m}{n}}$$

(3)

Q-SW:

$$\frac{\mu}{\mu_o} = \frac{1 + \left(\sqrt{k} - \sqrt{\frac{m}{n}}\right)\sqrt{\frac{m}{n}}\frac{\mu_m}{\mu_o}}{1 + \left(\sqrt{k} - \sqrt{\frac{m}{n}}\right)\sqrt{\frac{m}{n}}}$$

(4)

Q-LW:

$$\frac{\mu}{\mu_o} = \frac{1 + \left(\sqrt{k} - \frac{m}{n}\right)\frac{\mu_m}{\mu_o}}{1 + \left(\sqrt{k} - \frac{m}{n}\right)}$$

(5)

L-LW:

$$\frac{\mu}{\mu_o} = \frac{1 + \left(k - \frac{m}{n}\right)\frac{\mu_m}{\mu_o}}{1 + \left(k - \frac{m}{n}\right)}$$

(6)

L-NW:

$$\frac{\mu}{\mu_o} = \frac{1 + (k-1)\frac{m}{n}\frac{\mu_m}{\mu_o}}{1 + (k-1)\frac{m}{n}}$$

(7)

As shown in Eq. (1), the $\mu$ depends on the ratio of minority votes to majority votes $b/a$. Therefore, the number of $p$ does not appear in Eqs. (3) to (7). In other words, $\mu$ does not depend on the number $p$ but is determined by whether the votes are evenly distributed or cumulatively concentrated on the $k$-dimensional value coordinates.

## 3. Computation of Results

Figure 2 is a set of five graphs showing the results of Q-NW, Q-SW, Q-LW, L-LW, and L-NW using Eqs. (3) to (7). The horizontal axes of the five graphs represent the ratio $\mu_m/\mu_o$, and the vertical axes represent the ratio $\mu/\mu_o$. The $\mu/\mu_o$ ratio is calculated from $m/n$ shown in Eq. (2), $\mu_m < \mu_o$ represents the minimum possible value, and $\mu_m \geq \mu_o$ represents the maximum possible value.

In each graph in Fig. 2, the nine lines are drawn employing $3 \times 3$ parameters using the ratio of the standard deviation $\sigma_o$ to $\mu_o$ ($\sigma_o/\mu_o = 0.2, 0.5, 0.8$) and $k$ ($k = 1, 3, 5$). However, when $k = 1$, the numbers of votes of the majority and minority are equal ($a = b$), the minority is not weighted ($w_p = 1$), and $\mu/\mu_o$ is always 1, therefore the three lines ($\sigma_o/\mu_o = 0.2, 0.5, 0.8$ at $k = 1$) overlap. The diagonal light-gray line in each graph is a hypothetical case in which $\mu_m$ is completely in agreement with $\mu$, and that the opinions of the minority are more likely reflected as the calculation results become closer to this light-gray line.

Figure 3 is a graph comparing the calculation results of Q-NW, Q-SW, Q-LW, L-LW, and L-NW when $\sigma_o/\mu_o = 0.5$ and $k = 3$. The horizontal and vertical axes and diagonal light-gray line are the same as those in in Fig. 2. The horizontal light-gray line shows the case of $\mu/\mu_o = 1$ for reference.

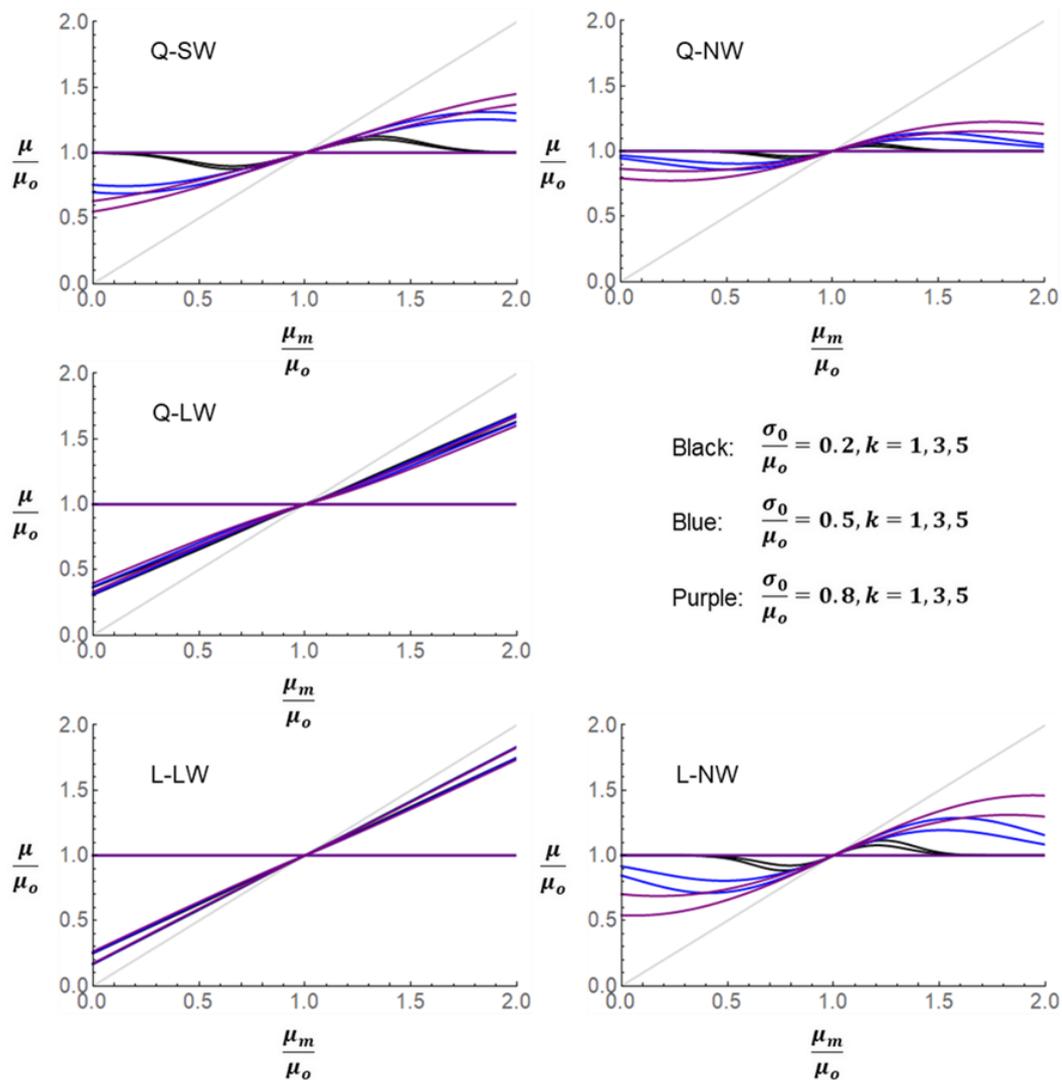

Figure 2. Calculation results of Q-NW, Q-SW, Q-LW, L-LW, and L-NW

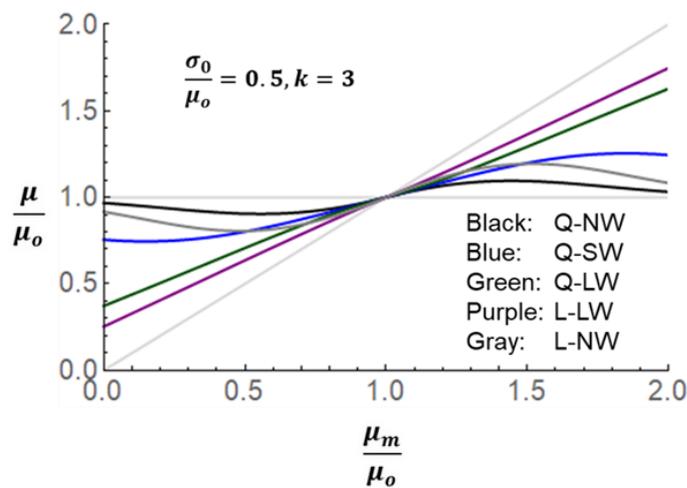

Figure 3. Comparison of Q-NW, Q-SW, Q-LW, L-LW, and L-NW

The calculation results in Figs. 2 and 3 generally show that the larger the $\sigma_o/\mu_o$ (the more the opinions of the majority spread in a certain one dimension), or the larger the $k$ (the more the opinions of the majority are distributed in many dimensions), the more the opinions of the minority are reflected. As a matter of course, Q-NW, Q-SW, Q-LW, L-LW, and L-NW have $\mu = \mu_m$ in the limit of $k \to \infty$ (approaching asymptotically to the diagonal light-gray lines in Figs. 2 and 3).

Looking at the calculation results of Q-NW and L-NW that do not give weighting to the minority's positionality shows that the $\mu/\mu_o$ of Q-NW depends on $b/a = \sqrt{k}$, and the $\mu/\mu_o$ of L-NW depends on $b/a = k$; therefore, the minority opinion is reflected in L-NW rather than in Q-NW.

When comparing Q-NW to Q-SW or L-NW to L-LW, minority opinions, are more likely to be reflected in the latter (i.e. Q-SW and L-LW) which give weighting to the positionality of the minority than in the former (Q-NW and L-NW), which does not give weighting to the positionality of the minority. As regards the weighted Q-SW, Q-LW, and L-LW, the minority opinions are more likely to be reflected in ascending order as follows: Q-SW, Q-LW, and L-LW as $b/a = \sqrt{k} \sim k$ and $w_p = \sqrt{m/n} \sim m/n$ increase.

In Q-NW, Q-SW, and L-NW, the shape of the graphs is a wave, so if $\mu_m$ is too far from $\mu_o$ ($\mu_m/\mu_o$ is too far from the centered value 1), it will be difficult to reflect the opinion of the minority. In Q-LW and L-LW, since the shape of the graphs is linear, there is a proportionate reflection of the opinions of the minority.

Note that Q-LW and L-LW have less dependence on $\sigma_o/\mu_o$ than the other methods. This means that the minority opinions are reflected regardless of the degree of concentration of the opinions of the majority. In Eqs. (5) and (6), for Q-LW and L-LW, $k$ and $m/n$ are in same parentheses. This is because when $m$ is small, $k$ becomes dominant in the parentheses, that is, $\sigma_o/\mu_o$ in Eq. (2) becomes ineffective.

In summary, by adopting the new aggregation methods (Q-SW, Q-LW, L-LW) that give weighting to the minority's positionality on cumulative voting for pluralistic values, they enable the social choices that reflect the opinions of the minority more than the usual cumulative voting method (L-NW) and the usual quadratic voting method (Q-NW). Of the three positionality-weighted aggregation methods, Q-LW and L-LW, which were weighted using a linear ratio of the majority to the minority, could impartially reflect the opinions of the minority even when the number of minority voters decreased.

The usual cumulative voting was devised to make it easier to reflect the opinion of the minority by allocating points, and the usual quadratic voting a variant of cumulative voting allocates the number of votes as the square root of points based on economic cost. In contrast with these, the new aggregation methods were weighted by the ratio of the majority and the minority $m/n$ or its square root $\sqrt{m/n}$, and the opinion of the minority was better reflected than when using the usual methods.

## 4. Discussion

These results obtained indicate that Sen and Gotoh's idea (Sen, 2002; Gotoh, 2015) of giving priority on ordinal ranking to unfortunate people in the position-conscious choice procedure is applied to the positionality of the minority on cardinal cumulative voting. In other words, by replacing the qualitative "position" information of Sen and Gotoh with the quantitative "positionality" information in the new aggregation methods, the axiomatic approach is transformed into a numerical approach. This makes it possible to incorporate minorities not only in axiomatic and formalistic academic fields, but also in voting scenes for actual social choices.

With the cardinal cumulative voting methods, the existence of a minority can be extracted from analyzing voting data. However, as described in Chapter 1, it is not possible to know the content of the social position such as a specific social hierarchy, victimhood, disability, race, and culture. We believe that the aggregation methods (including their further deformation methods) that should be adopted should be selected based on social issues and positions. For example, in the dimension related to economic value, Q-NW, originally derived from an economic perspective, will be adopted, and in the dimension related to environmental value, Q-SW or Q-LW will be adopted to balance economic value while considering the opinion of the minority. In the dimensions related to social and human values, it will be possible to adopt Q-LW or L-LW, which proportionately reflect the opinions of the minority.

One application of these methods to the real world is to the choice of operational measures with the three representative values, i.e., social values (regional economic circulation rate, that is, regional revitalization), environmental values (natural energy usage rate), and economic values (energy cost) in the local community (Hiroi, 2009; Edahiro, 2018). Another application is the choice of management methods regarding social value (transfer to descendants and others), environmental value (distribution of natural resources), and economic value (distribution of profits) in a mutual aid of a community (Najita, 2009). All these, are cases in which social choices are made from cardinal alternatives for pluralistic values. As information technology permeates throughout the real world, there will be more cases in which consensus building for not only ordinal and discrete alternatives but also pluralistic and cardinal alternatives will be required.

The consensus building is a process of seeking unanimous agreement (Susskind, 1999; Inohara, 2011). The proposed aggregation methods that give weighting to the minority's positionality can be used interactively in the process of consensus building among stakeholders with diverse values. This is as opposed to immediately determining the choice of the cardinal value based on the aggregation results. By visualizing the aggregation results and providing them to stakeholders in the process of consensus building, it is possible the majority and minority could deepen their discussions by taking each other's position into account.

For example, on an online interface for each voter, as same as online opinion surveys and quadratic voting, some cardinal sliders corresponding to the pluralistic value axes and some buttons for adding/deleting points of cumulative voting would be provided. Examples of visualization interfaces include the opinion position and number of minority voters, original distribution that is back-calculated to 1p1v (without weighting), original distributions of the majority and minority that decomposed the original distribution, strength of minority opinions that is the cumulative voting result, and aggregation result that gives weighting to the minority's positionality. Moreover, presenting information of not only one dimension but also other dimensions in the consensus-building process, facilitates mutual coordination and compromise between dimensions.

Although our calculation results were focused on a case in which there was one minority group in a certain dimension, when there is a conflict between minorities on the same dimension, there is no choice but to solve it in the consensus-building process by visualizing and recognizing the existence of each minority. Note that our results were calculated from the maximum number of minority voters in the original frequency distribution, as described above regarding Eq. (2). Social issues with extremely few minority voters should be evaluated as individual issues rather than addressing them through cumulative voting.

It should be noted that, similar to the usual cumulative voting method, the proposed aggregation methods will satisfy the conditions of anonymity, neutrality, and universality of domain. Similar to Sen and Gotoh's position-conscious choice procedure, the transition from the usual cumulative voting to the proposed aggregation methods means the addition of a condition that respects the minority opinions, and the potential for compatibility with the Pareto condition. We believe that the visualization of the above-mentioned information and its use in the consensus-building process will provide an opportunity for adjustment and reflection and robustly address the conspiracy of the majority and the irrationality of voters.

## 5. Conclusion

To build consensus on pluralistic values, we proposed aggregation methods that take into account the minority's positionality on cardinal cumulative voting. These methods incorporate the idea of the position-conscious procedures on an ordinal ranking scale proposed by Sen and Gotoh for unfortunate people into cardinal cumulative voting therefore easily reflecting the true opinions of voters. Our aggregation methods allow us to highlight (rather than bury) minority's votes, compared to the usual cumulative voting and the quadratic voting.

However, while minority opinions are more likely to be reflected in social choices than the conventional majority rule, there is concern that minority rule by selfish egoists will prevail. To address this issue, it would be useful to use different aggregation methods for different social issues and social positions, or to interactively visualize the aggregation contents during in the consensus building process between the majority and the minority. In the latter (i.e. visualization of aggregation contents), the selfish minority will be encouraged to compromise by visualizing its' selfishness against common goals and norms that have been agreed at the stage of participation in parliament or community.

Well-known consensus-building tools include the digital platform for participatory democracy (Decidim, 2021) and the online collective decision-making tool (Loomio, 2020). These tools include features such as proposals, assessments, meetings, discussions, bulletin boards, voting, and choices. The voting feature supports several voting methods such as majority voting, Borda voting, and cumulative voting, and voting results are displayed in bar graphs. If we add our new aggregation methods to such a conventional voting feature, as discussed in Chapter 4, not only will they display bar graphs, but also the position and number of minorities, and the distribution of majority and minority, etc. By doing so, we believe that the position-conscious choice procedure aimed at by Sen and Gotoh can be realized through a consensus building process supported by practical IT tools. With well-being, equality, and inclusive societies as set goals in the SDGs of United Nations (Sustainable Development Goals, 2015), and it seems that social choice that considers "positionality" will occupy an important position in the future.

With the further development of information technology, there will be advances in the integration of social systems and IT systems, represented by the cyber-physical system. Therefore, analysis and simulation of social issues based on big data will be conducted, and consensus building on cardinal choices will be necessary. We propose the use of such aggregation methods to make social choices for pluralistic values such as social, environmental, and economic.